\documentclass[twocolumn,preprintnumbers,amsmath,amssymb]{revtex4}
\usepackage{graphicx}% Include figure files
\usepackage{dcolumn}% Align table columns on decimal point
\usepackage{bm}% bold math
\begin{document}
\title{\bf Thermal properties of fluorinated graphene}
\author{S. K. Singh$^1$, S. Goverapet Srinivasan$^2$, M. Neek-Amal$^1$, S. Costamagna$^{1,3}$, Adri C. T. van Duin$^2$, and F. M. Peeters$^1$}

\affiliation{$^1$Universiteit Antwerpen, Department of Physics,
Groenenborgerlaan 171, BE-2020 Antwerpen, Belgium.\\$^2$ Department of Mechanical and Nuclear Engineering,
The Pennsylvania State University, University Park, Pennsylvania 16801, USA.\\ $^3$Facultad de
Ciencias Exactas Ingenier{\'\i}a y Agrimensura, Universidad Nacional
de Rosario and Instituto de F\'{\i}sica Rosario, Bv. 27 de Febrero
210 bis, 2000 Rosario, Argentina.}

\date{\today}

\begin{abstract}
Large scale atomistic simulations using the reactive force field
approach (ReaxFF) are implemented to investigate the
thermomechanical properties of fluorinated graphene (FG). A new set
of parameters for the reactive force field potential (ReaxFF)
optimized to reproduce key quantum mechanical properties of relevant
carbon-fluor cluster systems are presented. Molecular dynamics (MD)
simulations are used to investigate the thermal rippling behavior of
FG  and its mechanical properties and compare them with graphene
(GE), graphane (GA) and a sheet of BN. The mean square value of the
height fluctuations $\langle h^2\rangle$ and the height-height
correlation function $H(q)$ for different system sizes and
temperatures show that FG is an un-rippled system in contrast to the
thermal rippling behavior of graphene (GE). The effective Young's
modulus of a flake of fluorinated graphene is obtained to be 273 N/m
and 250 N/m for a flake of FG under uniaxial strain along arm-chair
and zig-zag direction, respectively.

\end{abstract}


%%%%%%%%%%%%%%%%%%%%%%%%%%%%%%%%%%%%%%%%%%%%%%%%%%%%%%%%%

\maketitle

%%%%%%%%%%%%%%%%%%%%%%%%%%%%%%%%%%%%%%%%%%%%%%%%%%%%%%%%%
\section{Introduction}
The fascinating properties of single layer graphene (GE)
have triggered a broad interest in the solid state
physics community~\cite{novo,Geim, Berger, Katsnelson, Sahin}.
 Despite its high electron mobility~\cite{castro-neto}, the zero band gap defies its employment
in nano transistors where it is desirable to have a large on-off
ratio between conducting and non-conducting states. A band gap can
be induced by the addition of adatoms, which changes locally the
hybridization of the carbon (C) atoms, but also modifies the electron
mean free-path affecting the electron transport properties. Hydrogen
(H) and fluor (F) are two well tested
candidates~\cite{elias,Nair,robinson,zboril} that leads to a large
band gap opening. Graphane (GA, hydrogenated graphene) and
fluorographene (FG) have been studied both experimentally and
theoretically~\cite{Cheng, Withers,leenaers1,samarakoon} to engineer
the band gap.

When H or F atoms are attached to the C atoms of GE, the bonds
transit from an {\it sp$^2$} to an {\it sp$^3$} hybridization, which
turns the conjugated graphitic C-C bonds into single C-C bonds. In the
fully covered cases both GA and FG are insulating
materials~\cite{elias,Nair} and the structure changes locally the
planar shape of GE into an angstrom scale out-of-plane buckled
shaped membrane~\cite{PRB77} known as {\textit {chair
configuration}}~\cite{sluiter,sofo}.

From its potential applications in nanotechnology, FG is a more favourable material than GA since the C-F bonds are energetically more stable
than the C-H bonds~\cite{leenaers1,PRB77,sofo,Sahin1}. Fluorographene  has a very large
temperature-dependent resistance and when the fluor content is
increased it induces large changes in the electron
transport~\cite{freddie1}. As in GE, it is expected that
temperature also affects strongly the lattice structure and
the mechanical properties of FG.

 According to the Mermin-Wagner
theorem,~\cite{mermin}, thermal excited ripples in two
dimensional-like materials (GE, bilayer GE, GA and FG) has to play an important role in the stability of the membrane. While in GE and bilayer GE
 the corrugations are well described within the theory
of two dimensional continuous membranes~\cite{bilayer,3}, for GA
instead, we recently found that the angstrom scale buckling of the
carbon layer of GA prevents the formation of intrinsic long
wavelength thermal ripples for
temperatures up to at least 900 $K$~\cite{seba2012}.

Since the C atom has a higher (lower) electronegativity than
 H (F), it will take (give) away charge from the  H (F) atom and  consequently transforms the resulting C-H and C-F covalent
bonds into sp$^3$ bonds. Therefore, it is expected that similar
rippling effects as in GA will occur in FG although the C-F bonds
are somewhat stronger than the C-H bonds. The latter is due to the larger amount of charge that is shifted
from C to F as compared to the one from H to C~\cite{leenaers1}.
However in order to simulate large scale FG samples an appropriate
force field is needed
 which describes the  true chemical bond in C-F. Indeed, the absent of such a suitable interatomic
  potential for C-F restricted  most of the recent studies to ab-initio calculations of their electronic properties
 using a small computational unit cell.

ReaxFF potential serves to describe both bond and non-bond interactions in solids. Recently, such potentials were parameterized and were well tested for different
kind of structures, e.g. hydrocarbons~\cite{van}, carbon
allotropes~\cite{Mueller}, etc. In this study we present a new set of
parameters for ReaxFF, appropriate for structures with C-F bonds.
Using molecular dynamics (MD) simulations over large scale samples we
 study the thermal corrugations of FG and compare the results
with those found for GA, GE and BN. We show that fully covered FG
follows the same trend as GA and does not develop long-wavelength
ripples or significant corrugation. The bending rigidity $\kappa$ of
FG is found to be larger than the one of GE, GA and BN. Furthermore,
$\kappa$ turns out to be temperature independent. Our results
indicate that long-wavelength ripples are instead present in partial
covered FG samples with a larger  amplitude as compared to GA.

The paper is organized as follows. In Sec. II, we introduce a new
set of parameters for the ReaxFF potential of the C-F covalent
bond. Then, in Sec. III using the introduced parameters, we analyze
the thermal rippling behavior. Here, we consider both fully and
partially covered graphene sheets by F atoms. All the results are
compared with those previously found for graphane. We also estimate
the effective Young's modulus of FG flakes. We conclude the paper in
Sec. IV.

%%%%%%%%%%%%%%%%%%%%%%%%%%%%%%%%%%%%%%%%%%%%%%%%%%%%%%%%%
\section{ReaxFF potential for Fluorographene}
\begin{figure}
\includegraphics[width=0.50\textwidth]{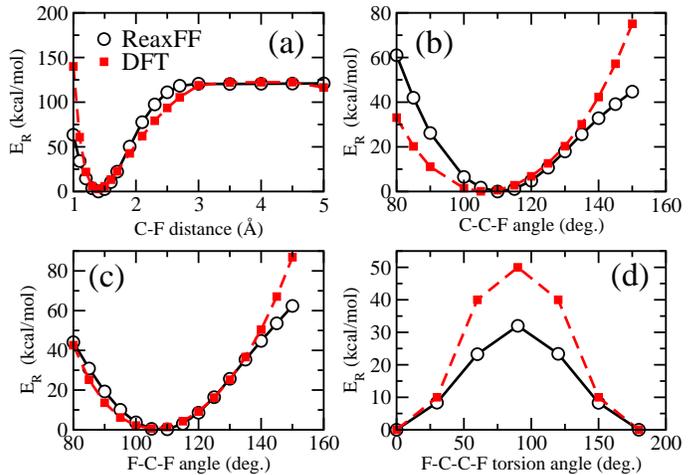}
\caption{(Color online) Comparison between DFT (solid squares) and
ReaxFF (open circles) results for: (a) C-F bond dissociation in
H$_{3}$C-CF$_{2}$-CH$_{3}$ (b) C-C-F angle bending in
H$_{3}$C-CF(CH$_{3}$)-CH$_{3}$, (c) F-C-F angle bending in
H$_{3}$C-CF$_{2}$-CH$_{3}$ ,and (d) F-C-C-F dihedral twisting in
F$_{2}$C=CF$_{2}$.} \label{ReaxFF2} \vspace{0.2cm}
\end{figure}

\begin{figure}[t]
\includegraphics[width=0.40\textwidth]{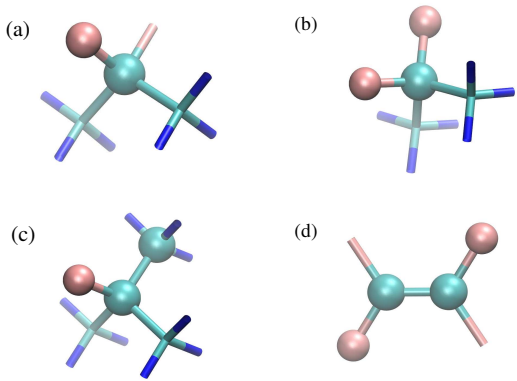}
\caption{(Color online) (Color online) The molecules used for
parameterizing the ReaxFF force field in this study. (a) C-F bond in
H$_3$C-CF$_2$-CH$_3$ (b) F-C-F angle in H$_3$C-CF$_2$-CH$_3$ (c)
C-C-F angle in H$_3$C-CF(CH$_3$)-CH$_3$, and (d) F-C-C-F dihedral
angle in F$_2$C=CF$_2$. The atoms constituting the geometric
parameters are represented by balls while the rest of the atoms are
represented by sticks. F atoms are colored brown, C atoms are
colored green and H atoms are colored blue. } \label{figReaxFF}
\vspace{0.2cm}
\end{figure}

ReaxFF~\cite{van} is a general bond-order dependent potential that uses a relationship between bond distance and bond
 order on the one hand and a relationship between bond order and bond energy on the other hand to describe bond formation
  and dissociation. Many body interactions such as the valence angle and torsional interactions are formulated as function
  of bond order so that their energy contributions vanish smoothly upon bond dissociation. Non bonded interactions, namely
  Coulomb and van der Waals interactions, are calculated between every pair of atoms irrespective of their connectivity.
Excessively close range interactions are avoided by shielding.
ReaxFF uses the  geometry dependent charge calculation scheme (EEM
scheme) of Mortier~ \emph{et al}~\cite{mortier}.  The system energy
in ReaxFF consists of a sum of terms:
\begin{eqnarray}\nonumber
E_{sys} = E_{bond} + E_{under} + E_{over} + E_{lp} + E_{val} + E_{pen} +
&&\\ \nonumber
E_{tors} + E_{conj} + E_{vdWaals} + E_{Coulomb}.
\end{eqnarray}
A detailed description of each of these terms and their functional
forms can be found in the original work~\cite{van}. The reactive
force field for C/F containing systems was developed by
parameterizing the potential against DFT data obtained at the
B3LYP/6-31g** level of theory (which is implemented in
Schrodinger~\cite{S11} which is an electronic structure packages)
for various quantities such as fluorine and carbon atom charges in
H$_{3}$C$-$CF$_{2}-$CH$_{3}$, C$-$F and C$-$C bond lengths in
H$_{3}$C$-$CF$_{2}-$CH$_{3}$ and H$_{3}$C$-$CF(CH$_{3}$)$-$CH$_{3}$,
F$-$F bond length in the F$_{2}$ molecule, potential energy curve
 for C-F bond dissociation in H$_{3}$C$-$CF$_{2}-$CH$_{3}$, F-C-F angle bending in H$_{3}$C$-$CF$_{2}-$CH$_{3}$, C-C-F angle bending in H$_{3}$C$-$CF$_{2}-$CH$_{3}$
 and F-C-C-F dihedral twisting in F$_{2}$C$=$CF$_{2}$ along with various chemical reactions involving fluoroalkanes and fluoroalkenes. The results of the force field
  training are presented in Figs.~\ref{ReaxFF2}(a)-(d) and in Table~\ref{TablerexFF}. Fig.~\ref{figReaxFF} depicts the geometrical quantities relevant to Figs.~\ref{ReaxFF2}(a)-(d).
   It can be seen from Table~\ref{TablerexFF} that ReaxFF reproduces closely the DFT based equilibrium geometries for various compounds. ReaxFF predicts F$_{2}$ dissociation
    energy of 36.6 kcal/mol, in very good agreement with the DFT value of 37 kcal/mol. Fig.~\ref{ReaxFF2}(a) shows that ReaxFF based potential energy curve for the C-F bond
    dissociation in H$_{3}$C$-$CF$_{2}-$CH$_{3}$ closely follows the DFT based potential energy curve. ReaxFF is able to predict very precisely the equilibrium C-F bond length
    (see Table~\ref{TablerexFF}) and the C-F bond dissociation energy. Similarly the force field can closely reproduce the DFT based potential energy curve and the equilibrium
    geometry (see Table~\ref{TablerexFF}) for C-C-F angle bending and the C-F-C angle bending as shown in Figs.~\ref{ReaxFF2}(b)-(c). Figure~\ref{ReaxFF2}(d) shows the variation
    of the potential energy upon F-C-C-F dihedral angle twisting. Though ReaxFF predicts the correct trend, the torsional rotation barrier in ReaxFF is around 18 kcal/mol lower than
    that predicted by DFT. Overall, the ReaxFF force field for C/F systems can closely reproduce the DFT based energies and geometries for a number of molecules and reactions.
    This force field will now be employed in  large scale fully reactive molecular dynamics simulation of C/F containing systems.

In the next section we study the thermal structural fluctuations and mechanical
properties of a single layer of FG
using large scale atomistic simulations employing the presented
ReaxFF parameters. These parameters
were implemented in the large-scale atomic/molecular massively parallel
simulator package LAMMPS~\cite{lammps,Plimpton}.

\begin{table*}
\caption{Comparison of equilibrium geometrical parameters between ReaxFF and DFT.}\label{TablerexFF}
\begin{tabular}{ c c  c  }
\hline\hline

&&\\
&DFT&ReaxFF\\
\hline
&&\\
F$_2$ bond length                          &  1.43\AA   &     1.4012\AA \\
C-F bond length in H$_3$C-CF$_2$-CH$_3$    &  1.3841\AA &     1.4057\AA \\
C-F bond length in H$_3$C-CF(CH$_3$)-CH$_3$&  1.3841\AA &     1.4158\AA\\
Non-bonding C-F distance in CF$_2$ dimer   &  2.00 \AA  &     2.4471\AA\\
F-C-F angle in H$_3$C-CF$_2$-CH$_3$        &105.65$^{o}$&  107.2197 $^{o}$\\
C-C-F angle in H$_3$C-CF(CH$_3$)-CH$_3$    &106.2 $^{o}$&  109.9625 $^{o}$\\
\hline\hline
\end{tabular}
  \centering

\end{table*}

\section{Results}
\subsection{Thermal rippling behavior of FG}

\begin{figure}[t]
\includegraphics[width=0.45\textwidth]{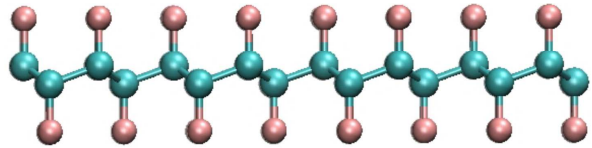}
\caption{(Color online) (a) Side view of the buckled structure, known as chair configuration,
of fully fluorinated graphene. The averaged bond-angle, and C-C and C-F distances,
are respectively 109.5$^{\circ}$, $d_{C-C}=$1.58 $\AA$ and $d_{C-F}=$1.41 $\AA$ at room temperature. } \label{fig1} \vspace{0.2cm}
\end{figure}

In order to study the rippling behavior of FG we considered a square
shaped computational unit cell of FG with both armchair and zigzag edges in the $x$
and $y$ directions.
Partial fluor contents of 10$\%$, 50$\%$,
70$\%$, 90$\%$ and the fully covered 100$\%$ case (with a total number of  $N=17280$ atoms)
were studied. In our simulation we employed the NPT ensemble with $P$=0 using the
Nos\'{e}-Hoover thermostat and varied the temperature from 10\,K to 900\,K.
Figure~\ref{fig1} shows the obtained buckled shape of fully fluorinated sample after
relaxation which is in agreement with recent DFT results~\cite{leenaers1}.

One would expect that the thermal excited ripples in FG can be described by membrane theory for a 2D continuous
membrane~\cite{Nelson}. This theory,
described in a series of related  works~\cite{seba2012,GE2d,seba1,roldan},
is supposed to be universal and independent of the atomic scale details of the membrane.
The main predictions of this theory are as follows.
Let $h$ be the out-of plane displacement of a given atom of a sheet, then
the Fourier transform of the height-height correlation function is in the harmonic approximation given by
\begin{eqnarray}
%%\begin{equation}
H(q)=\langle|h(q)|^2\rangle=\frac{N k_B T}{\kappa S_0 q^{4}}~~,
\label{hq1}
%\end{equation}
\end{eqnarray}
where $q$ is the wave-vector, $N$ is the number of atoms, $S_0$ is the surface
area per atom, $ \kappa $ is the bending rigidity of the membrane,
and $k_B$ is the Boltzmann constant.

In the large wavelength limit, anharmonic couplings between bending and stretching modes are important resulting in a renormalization of the
$q$-dependent behavior
\begin{eqnarray}
%\begin{equation}
H(q)=\frac{N k_B T}{\kappa S_0 q^{4-\eta}}
\label{hq2}
%\end{equation}
\end{eqnarray}
where $\eta$ is an universal scaling exponent which is about
$\approx 0.85$~\cite{ledousal,LCBOPII,fas1}.

In order to compare our results for FG with other two dimensional
materials, we used a modified Tersoff potential (which is an
ordinary defined potential in the LAMMPS package~\cite{Plimpton})
according to the parameters proposed by Sevik~\emph{et al} for the
h-BN sheet~[\onlinecite{Sevik}]. To simulate GE and GA we have used
the AIREBO potential~\cite{AIREBO} which is suitable for simulating
hydrocarbons.

Recently, we found that in GA, $H(q)$ acquires a strong
renormalization for small wave-vectors $q$ and the layer  remains
almost flat even for  temperatures as high as 900 K~\cite{seba2012}.
Here we will analyze the thermal rippling behavior of FG and compare
it with GA. A comparison with GE and BN single layers which behave
as 2D membranes~\cite{GE2d,seba3} will also be presented. $H(q)$ for
FG was calculated following the steps described in our previous
work~\cite{seba2012}.

\begin{figure}[t]
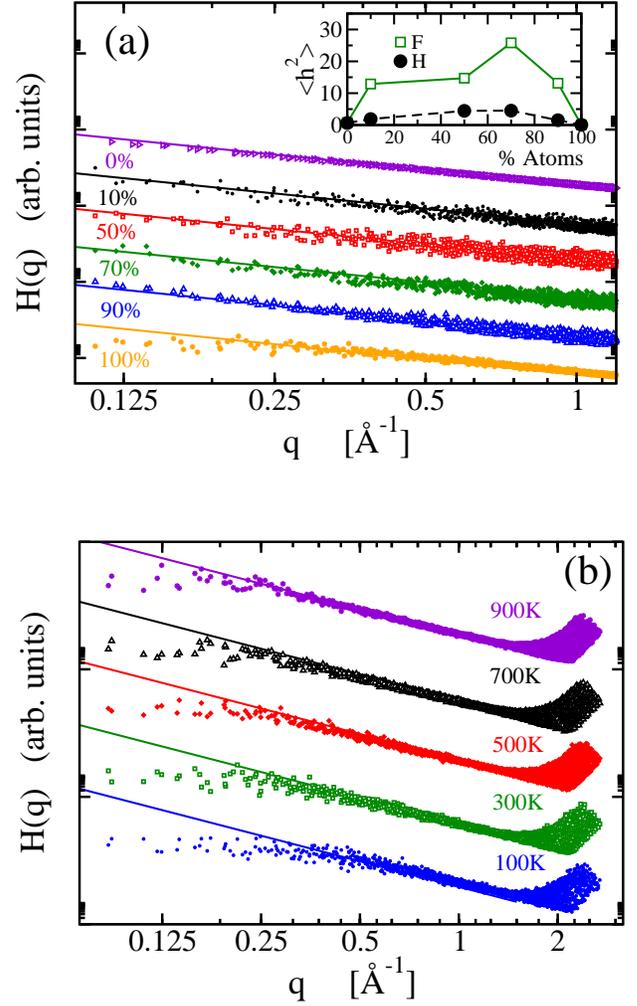

\includegraphics[width=0.45\textwidth]{Hq_partial_flour.eps}
\vspace{1cm}

\includegraphics[width=0.45\textwidth]{Hq_flour-temp.eps}
%\hspace{0.2cm}
%\includegraphics[width=0.3\textwidth]{kappa_FG.eps}
\caption{(Color online) (a) Log-log plot of $H(q)$ for different coverage of F atoms
 at T=300 K. The solid lines show the harmonic
$q^{-4}$ behavior valid in the limit of large $q$ values. Note the
strong deviation, starting roughly at $q^* \approx 0.2 \AA^{-1}$ in
the limit of long wave-lengths, for the case of fully fluorinated
graphene. The variation of $\langle h^2 \rangle$ is shown in the
inset.  (b) $H(q)$ for fully fluorinated FG at different temperatures.
%Variation of (c)  $\langle h^2 \rangle$, and (d) $\kappa$ against temperature for FG (open squares), GA (filled squared),  GE (open circles), and BN (stars).
} \label{fig12}
\end{figure}

Starting from a pure GE sheet, the variation of the height-height correlation function $H(q)$
at room temperature for different partial fluor contents is shown in Fig.~\ref{fig12} (a).
The curves were shifted for a better comparison.
We found that while for 10 to 90$\%$ coverage, $H(q)$ follows Eq.~(\ref{hq2}) up to small $q-$values which is similar to the case of GE~\cite{seba1}. But,
 for fully FG at $q^* \approx 0.2 \AA^{-1}$, $H(q)$ deviates
from the harmonic law (solid line) and approaches roughly a constant value similar to was previously found for GA~\cite{seba2012}.
In the inset of Fig.~\ref{fig12}(a) we show the square average of the out-of-plane fluctuations $\langle h^2 \rangle$ at 300~K.
Notice that the out of plane fluctuations for partially covered samples are considerably larger for FG than for GA.
The temperature dependence of $H(q)$ for fully fluorinated graphene is shown in Fig.~\ref{fig12}(b).
Irrespective of temperature,  the short
wave-length limit of  $H(q)$ tends always approximately to a constant value. The characteristic $q$-value where $H(q)$ deviates
from the harmonic approximation result decreases with increasing temperature.

\begin{figure}[t]
\includegraphics[width=0.49\textwidth]{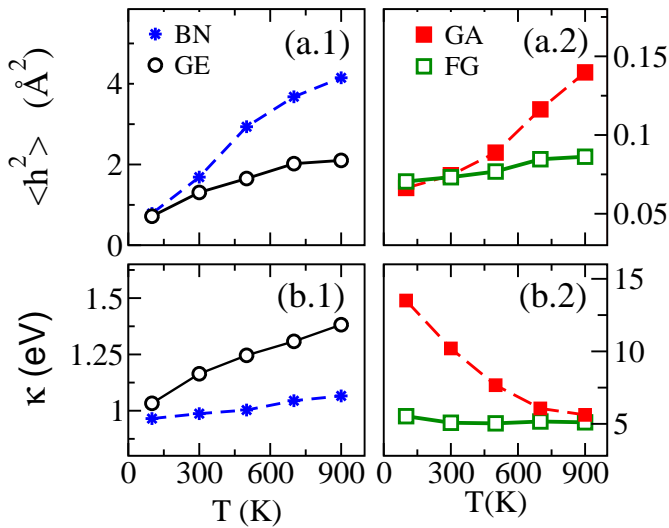}
\caption{(Color online) Variation of (a)  $\langle h^2 \rangle$,
and (b) $\kappa$ against temperature for FG (open squares), GA (filled squared),  GE (open circles), and BN (stars).}
\label{fig13}
\end{figure}

The renormalization of $H(q)$ for long wavelengths  indicates the
suppression of large out-of-plane height fluctuations. In
Fig.~\ref{fig13}(a) we compare the behavior of $\langle h^2 \rangle$
against temperature for GE, BN (panel (a.1)), FG and GA (panel
(a.2)). Notice that  $\langle h^2 \rangle$ increases from
$0.7~\AA^2$ up to $4~\AA^2$ in BN and from  $0.7~\AA^2$ up to
$2~\AA^2$ in GE when temperature is varied from  $10~K$ up to 900~K.
Due to the absence of long wave-length ripples, $\langle h^2
\rangle$ remains approximately constant for GA and FG, and the
variations are smaller than those for BN and GE, over the same
temperature range. The temperature dependence of the bending
rigidity $\kappa$, computed from the harmonic part of H(q) is shown
in Fig.~\ref{fig12}(b). Note that the larger magnitude for GA and FG
is a consequence  of the smaller corrugations present in these
materials. We also find the opposite temperature dependence for BN
and GE when compared with GA and FG. In this sense GE and BN behave
anomalously. The corresponding bending rigidity and $\langle
h^2 \rangle$ at room temperature for GE, GA and FG are listed in
Table II.

\begin{table}
\begin{tabular}{ c c  c  c| c c c}
\hline\hline
&&\\
&&AIREBO&&&&ReaxFF\\

&&$\kappa(eV)$&$\langle h^2\rangle$(\AA$^2$)~&&$\kappa(eV)$&$\langle h^2\rangle$(\AA$^2$)~\\
\hline

GE    &  &1.165 &1.307&&1.16&0.627 \\
GA    &  &10.19  &0.070&&7.26&0.074 \\
GF&   & -- &-- &&5.07&0.073 \\
\hline\hline
\end{tabular}
  \centering
\caption\textbf{{Comparison of AIREBO and ReaxFF for the bending
rigidity and $\langle h^2\rangle$ for GE, GA and FG at
300\,K.}}\label{TablerexFF}
\end{table}

Density functional calculations for band structure of FG (and GA)~\cite{APLHar} shows that the acoustic
out-of-plane modes (ZA) in FG (and GA) are different from that of
 GE. The most important difference from GE is the decoupled optical and
acoustic bands in FG and GA close to the  $\Gamma$ point.
 The light H atom contributes to the highest phonon frequencies which is not the
case for F atoms. It is also seen that the  ZA modes close to the  $\Gamma$
point for FG (and GA) are not well fitted by a quadratic function in contrast
to the GE case. This is clear indication of anharmonicity. In summary,
the atomistic details of the structure of FG is more complicated and therefore
more details of this structure should be included in any continuum
theory.

Before ending this section, note that as we discussed in our
previous work~\cite{seba2012}, the scaling with the system size
present in GE is no longer valid for  FG and GA ($\langle h^2 \rangle$
in FG and GA is almost constant irrespective to the system size).
The lower wavelengths ($q$) adopted for the calculation of $H(q)$
are equal to $q_{x-min}=\frac{2\pi}{l_{x}}$ and
$q_{y-min}=\frac{2\pi}{l_{y}}$ and represent the 'computational' cut-offs of
possible large wavelength ripples where $l_x$ and $l_y$ are the
dimension of the system. Notice that deviation from
the harmonic behavior takes place at larger value of $q$ and hence
this effect can not be a finite size effect and is instead an
intrinsic phenomenon of the material.

%%%%%%%%%%%%%%%%%%%%%%%%%%%%%%%%%%%%%%%%%%%%%%%%%%%%%%%%%
\subsection{Effective Young's modulus}
 In order to study the mechanical stiffness we consider an FG flake with dimension
$l_x\times l_y=15\times15$\,nm$^2$. %We applied  uniaxial stress on
%two opposite longitudinal ends and made sure to that the system was
%in thermal equilibrium while applying an infinitesimal velocity
%(0.005\,nm/ps) at the ends of the flake. More details of the method
%employed can be found in our previous works~\cite{neekprb82}.
Before the stretching process, the sample is equilibrated for 5 ps
(i.e. 50,000 time steps). Stretching direction is always along $x$
and the uniaxial strain is applied within the NPT
ensemble~\cite{NanoTech} where the pressure is slowly increased,
i.e. 2\,GPas/ps. In this section the lateral edges (in the $y$
direction) were taken as the arm-chair direction having both free
(FBC) and periodic boundary conditions (PBC).
 We kept temperature fixed at T=10\,K.

The total strain energy per atom of the strained flake can be written as
a function of the imposed strain ($\epsilon$)
\begin{equation}
E_T(l_y,\epsilon)=E_0+\frac{S_0}{l_y}\gamma(\epsilon)+\frac{S_0}{2}Y\epsilon^2
\label{Estrain}
\end{equation}
where $E_0$ is the energy of the infinite planar undeformed flake,
$\gamma(\epsilon)$ is the excess edge energy, and $Y$ is  Young's
modulus ($Y$) of the flake.

For nano-ribbons with no lateral edges we have $\gamma=0$ (assuming
that the longitudinal edges which are fixed make no contribution).
This is due to the fact that free edges increase the energy due to
 buckling and bond-order changes~\cite{Lu2010}. Recently Lu \emph{et al} used
  the Brenner potential~\cite{brenner} in
molecular dynamics simulations and studied the excess edge energy of
graphene nano-ribbons as a function of width and
chirality~\cite{Lu2010}. Our systems are different from those of
Ref.~\cite{Lu2010}. In contrast to Ref.~\cite{Lu2010} we are not
interested in effects due to the edge energy effect and the size
dependence. We rewrite Eq.~(\ref{Estrain}) in the following as an
effective Young's modulus which qualitatively gives a good
description of the mechanical stiffness of all the examined 2D
materials. Nevertheless our results are in qualitative agreement
with those reported by Lu~\emph{at al}, i.e. increasing of total
energy for the FBC case as compared to a nano-ribbon. Assuming a
quadratic relation for $\gamma(\epsilon)=\frac{l_y}{2}\epsilon^2$
valid for small $\epsilon$, the simplest method to estimate Young's
modulus is by fitting the quadratic function to the total energy
(per area):
\begin{equation}
E_T=E_0+\frac{1}{2}Y_{eff}\epsilon^2, \label{Eeff}
\end{equation}
where $Y_{eff}$ is the effective Young's modulus of the system.
  Using aforementioned fitting process we found $Y_{eff}$ respectively for a flake with arm-chair and zig-zag FG, to be 273 N/m and
  250  N/m. Notice that the experimental result is 100 N/m for not perfect FG~\cite{Nair} while the DFT result is 250
  N/m~\cite{Sahin1}. The latter disagreement between theory and
  experiment may be explained due to the fact that in experimental samples the fluor-to-carbon ratio is larger than
  unity, i.e.~1.1~\cite{Nair}, because of the presence of defects.
 Such defects become active
  regions which can adsorb the free F (and even H) atoms. Therefore, in
  the defected parts more F atoms will be found which is responsible for
  a F/C ratio larger than one.

\begin{figure}
\includegraphics[width=0.5\textwidth]{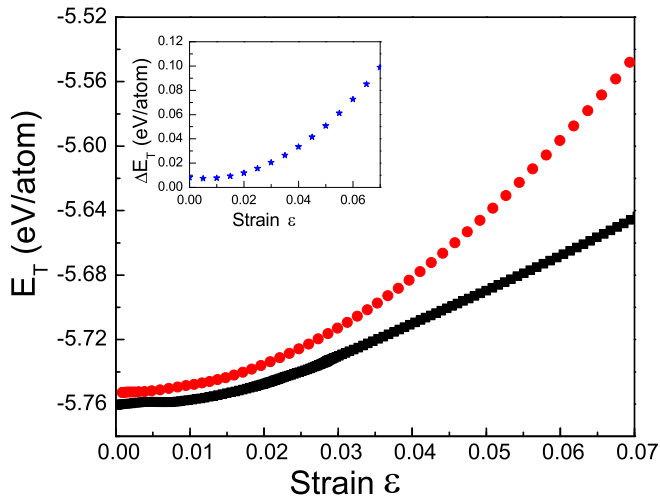}
\caption{(Color online) (a) Variation of total energy against
uniaxial strain for FG  subjected to free boundary condition (FBC)
and periodic boundary condition (PBC) for the lateral edges, i.e.
the dashed and solid curves respectively. The inset shows the
difference between the two energy curves. }
\label{figE}
\end{figure}

In order to understand the effect of the different boundary conditions, we depict in
Fig.~\ref{figE} the variation of $E_T$ per atom with
$\epsilon$ for flakes with both FBC (dashed lines) and PBC, i.e.
nano-ribbon (solid lines). It is seen that for flakes with FBC
the free edges result in an increase of the energy. The inset shows
the difference between two curves, i.e. $\Delta E_T=E_{FBC}-E_{PBC}$
which is positive. Because the free boundaries have many
dangling bonds which  are not saturated by F atoms it results in
extra energy. This can also occur in other systems, e.g. graphene~\cite{PRL2008}. Notice that
for the studied low temperature here, i.e.  T=10\,K we do not expect that
bond reconstruction at the edges is important. Notice that even by saturating all
the bonds by F, still the change
 in the bond order term in ReaxFF (due to different chemical environment of the boundary atoms) results in higher energy as compared to  PBC.

Furthermore, both FBC and
PBC results exhibit a quadratic behavior which is an indication of the harmonic regime for the
 applied uniaxial stain. As is clear from the inset of
 Fig.~\ref{figE}, the difference between the two curves  is increasing
 with applied strain. This is due to the deviation from equilibrium for the C-F bonds, C-C-F (F-C-F) bond angles,
 and the dihedral angles (F-C-C-F torsion angle)  of the free edge atoms.
 The larger the strain (and the larger the length of ribbon), the larger the deviation from
 equilibrium for the bonds and the angles. In the PBC case there is no such edge effect but nevertheless
 because of the fixing of the  two longitudinal ends (the edges which are under uniaxial
 stress) the energy variation of the PBC system should be different from
 that of an infinite FG which is periodic in both directions while it is
under tension from the arm-chair direction. The fixed longitudinal ends
do not have any effect in our results because both FBC and PBC have
the same contributions. 

%

%%%%%%%%%%%%%%%%%%%%%%%%%%%%%%%%%%%%%%%%%%%%%%%%%%%%%%%%%

\section{Conclusions}
We provided a new set of parameters for the ReaxFF potential for
 the C-F covalent bond and tested it on various molecules. Subsequently, molecular dynamics simulations were used to investigate
the thermal rippling behavior and the mechanical response of
fluorographene (FG) under  uniaxial stress. The obtained results are
compared with those for graphene (GE), graphane (GA) and hexagonal
boron nitride sheet (BN). We found that fluorographene remains a
flat sheet similar to graphane  even at high temperature, i.e.
up to 900 K. The bending rigidity of FG is found to be independent
of temperature and its Young's modulus is in good agreement with
experiment.

{\textit{Acknowledgments}}. \label{agradecimientos}  This
work is supported  by the ESF-Eurographene project CONGRAN, the
Flemish Science Foundation (FWO-Vl) and the Methusalem Foundation of the Flemish Government. SGS and ACTvD  acknowledge
support by the Air Force Office of Scientific Research (AFOSR) under
Grant FA9550-10-1-0563.

%%%%%%%%%%%%%%%%%%%%%%%%%%%%%%%%%%%%%%%%%%%%%%%%%%%%%%%%%%%

%%%%%%%%%%%%%%%%%%%%%%%%%%%%%%%%%%%%%%%%%%%%%%%%%%%%%%%%%%%

\end{document}